\documentclass[12pt,preprint]{aastex}
\begin{document}
\title{UPDATED PHOTOMETRY AND ORBITAL PERIOD ANALYSIS FOR THE POLAR \objectname{AM HERCULIS} ON THE UPPER EDGE OF THE PERIOD GAP}
\author{Zhibin Dai\altaffilmark{1},\email{zhibin\_dai@ynao.ac.cn} Shengbang Qian\altaffilmark{1} and Linjia Li\altaffilmark{2}}
\affil{Yunnan Observatories, Chinese Academy of Sciences (CAS), P. O. Box 110, 650011 Kunming, P. R. China.}
\altaffiltext{1}{Key Laboratory for the Structure and Evolution of Celestial Objects, Yunnan Observatories, Chinese Academy of Sciences (CAS), China.}
\altaffiltext{2}{University of Chinese Academy of Sciences, Beijing, P.R. China.}

\begin{abstract}
Twenty-one new optical light curves, including five curves obtained in 2009 and sixteen curves detected from the AAVSO International Database spanning from 1977 to 2011, demonstrate 16 new primary minimum light times in the high state. Furthermore, seven newly found low-state transient events from 2006 to 2009 were discovered, consisting of five Gaussian-shaped events and two events with an exponential form with decay timescales of $<$0.005 days; these timescales are one order of magnitude shorter than those of previous X-ray flare events. In the state transition, two special events were detected: a "disrupted event" with an amplitude of $\sim$ 2 mag and a duration of $\sim$ 72 minutes and continuing R-band twin events larger than all known R-band flares detected in M-type red dwarfs. All 45 available high-state data points spanning over 35 yr were used to construct an updated O-C diagram of \objectname{AM Herculis}, which clearly shows a significant sine-like variation with a period of 12-15 yr and an amplitude of 6-9 minutes. Using the inspected physical parameters of the donor star, the secular variation in the O-C diagram cannot be interpreted by any decided angular momentum loss mechanism, but can satisfy the condition $\tau_{\dot{\rm M}_{2}}\simeq\tau_{\rm KH}>>\tau_{\dot{\rm R}_{\rm 2}}$, which is required by numerical calculations of the secular evolution of cataclysmic variables. In order to explain the prominent periodic modulation, three plausible mechanisms¡ªspot motion, the light travel-time effect, and magnetic active cycles¡ªare discussed in detail.
\end{abstract}

\section{Introduction}

Polars, a subtype of magnetic cataclysmic variables (MCVs), play an important role in the study of interactions between two components. Since the accretion disk has been broken up by the primary, highly magnetic white dwarf, mass transfer may be seen more clearly for polars \citep[see][chap.6]{war95}. \citet{tap77} gives the magnetic field intensity of the white dwarf \object{AM Herculis}, $B_{\rm WD}\sim200$ MG, which is apparently higher than the 30 MG and 14 MG values reported by \citet{sch83} and \citet{wic85}, respectively. More recent papers have proposed that $B_{\rm WD}$ should be $\sim14$ MG \citep[e.g.][]{bai91,cam08}, in good agreement with
the value reported previously by \citet{wic85}. \objectname{AM Herculis} is the prototypical polar star that has been monitored extensively in optical bands \citep[e.g.][]{szk77,you79,cro81,hut02,kafa05,ter10}.
A typical feature in its optical flux is two distinct photometric
states: a high state with V $\sim$ 13.5 mag and a low state with
V $\sim$ 15.5 mag. This phenomenon in polars is commonly
attributed to variations in the mass transfer rate from the
secondary star. The high state is regarded as the normal accreting
state with an average mass accretion rate of $\sim10^{-10}$ M$_{\odot}$yr$^{-1}$ \citep{gan01,gan06}. As for the low state, \citet{liv94} proposed a starspot model in which the temporary cessation of mass transfer through the inner Lagrange (L1) point because of the migration of the starspot on the surface of the mass-donor star presumably causes the violent drop in the accretion luminosity. Then, the accretion rate would decrease to an extremely low level of $\leq$10$^{-12}$ M$_{\odot}$yr$^{-1}$, which may result
from the stellar wind of the late-type secondary star \citep{hes00,gan06}. Except for the starspot model, \citet{wu08} proposed that the variations in the magnetic field configuration of the binary system can also result in the alternation of the two states. At the present time, it is therefore still an open question as to the mechanism of high/low states in polars \citep[e.g.][]{kafa05,kaf09}.

The observed periodic modulations in the optical light curves of \objectname{AM Herculis} are thought to be the consequence of the self-eclipse of the accretion regions or part of the heated material surrounding white dwarf \citep{bai81,cro81,maz86,gan01}. The light minimum in a typical high-state light curve of \objectname{AM Herculis} has a broad duration of $\sim0.6 P_{\rm orb}$ \citep[e.g.][]{szk77,ols77,bai81}. Considering that the cyclotron emission near the polar(s) of white dwarfs may be a dominant source of the modulations in the V- and R-bands, \citet{you79} pointed out that the self-eclipse feature in photometric light curves has a stable modulation period, which is almost identical to the orbital period, in spite of the cycle-to-cycle variations \citep{szk80,gan01}. Based on this stability, many light minimum times from the ultraviolet to the infrared in both luminosity states have been obtained \citep[e.g.][]{szk77,pria78,maz86,hut02,kal08}. In addition, there are many ephemerides representing different periodic modulation signals. For example, the magnetic ephemeris is defined by the periodic time point when the sign of the circular polarization crosses zero \citep{tap77,hei88}. In addition, all optical light minimum times obey the original optical photometric ephemeris derived $\sim35$ yr ago \citep{szk77}; not all the times of the minima depend strongly on wavelength \citep{hut02}. Initially, \citet{maz86} carried out an O-C analysis and found that the variations in the O-C diagram are possibly caused by the oscillations of the white dwarf magnetic pole relative to the binary system line of centers \citep{cro81,cam85}. However, the following O-C diagrams for the high-state data alone indicated that its orbital period is almost constant \citep{kafa05}. More recently, \citet{kal08} argued that the ascending O-C curve, which is totally different from the previous O-C analyses, is a result of mass transfer from the low-mass red dwarf to the massive white dwarf. This assertion may imply that \objectname{AM Herculis} would become a detached binary system in the future. So far, there is not a common conclusion for the cause of the O-Cs of the primary light minima in the high state.

In this paper, the optical photometry of \objectname{AM Herculis} in low/high states from our observations and from the American Association of Variable Star Observers (AAVSO) is presented in Section 2. All collected data extend the time baseline of the O-C diagram to 35 yr. In Section 3, \citet{kafa05} photometric ephemeris of \objectname{AM Herculis} is corrected based on our new photometric data. Then, we analyzed the updated O-C diagram for all primary high-state minima in Section 4. Section 5 includes a discussion of the seven transient events and the possible explanations for the new O-C diagram.

\section{Data ACQUISITION/REDUCTION}

\subsection{Photometeries}

An \objectname{AM Herculis} photometric campaign was carried out from April to October 2009 using the three ground-based optical telescopes in China: the 60 cm and 1.0 m Ritchey-Chretien-Coude reflecting telescopes at Yunnan Astrophysical Observatory and the 85 cm Cassegrain telescope at Xinglong Observing Station of National Astronomical Observatories. The details of the photometry for \objectname{AM Herculis} can be seen in Table 1. All images were reduced using PHOT (measuring magnitudes for a list of stars) in the aperture photometry package of IRAF. We chose two nearby photometric standards for \objectname{AM Herculis}, AM Her-2 and AM Her-15, to be the comparison and the check star in our observations, respectively \citep{hen95}. The magnitude differences between the comparison and the check star are represented by the dotted lines in Figures 1(l), (m), (n). Figures 2(e) and (f) suggest that our photometry was carried out in good seeing and that the observed variations in the light curves are genuine and believable.

The three primary CCD minimum times were derived from the high-state differential light curves shown in Figures 1(l), (m), and (n) using a least-squares fitting method to a parabolic curve. The fitting errors and the time resolution of the CCD observations were combined to estimate the uncertainties of the minimum times. The other two low-state light curves are shown in Figures 2(e) and (f) for the analysis of transient events in Section 4.1. Additionally, the two white-light light curves, i.e., Figures 1(n) and 2(e), are marked by the letter N in the top-right corner of plots. Besides the positive color indexes, i.e., B-V $\sim$ 0.5 and V-R $\sim$ 0.6 \citep[e.g.][]{bai77,prib78,bai81}, and the flat B light curves \citep[e.g.][]{ols77,bai77,maz86,gan01}, the total spectral energy distribution derived by \citet{pric78} and \citet{bai81} demonstrate that the unfiltered photometries of \objectname{AM Herculis} are mainly dominated by radiation in the long-wavelength band and beyond.

\subsection{Light Minimum Times from the Literature}

The photometric ephemeris in the far-ultraviolet band derived by \citet{hut02} is different from that derived completely in the visual band \citep{szk77,kafa05}. The offset of light minimum times between the FUV/X-ray and optical bands is almost up to half an orbital period \citep[e.g.][]{hea77,maz86,hut02}. Therefore, \citet{kal08} were not justified in combining the average of the light minimum times in the FUV with those in the optical for their O-C analysis. In this paper, all 29 available photometric data points in HJD, including the data extracted from the literature, are listed in Table 2. Since \objectname{AM Herculis} has two distinct photometric states corresponding to different accretion configurations, we reexamined the photometric states of all 29 timing datasets.

Note that the three light minimum times obtained by \citet{kal08} present significant deviations of $\sim0.5$ $P_{\rm orb}$ from the photometric ephemerides derived by \citet{szk77} and \citet{kafa05}. Although these light minimum times may be the times of secondary minima, \citet{kal08} never made any special mention of them. Hence, these times cannot be included in our O-C analysis because of the instability of secondary minima \citep{bai77,szk80,kafa05}.

\subsection{Data from AAVSO}

The AAVSO is famous for its abundant information and data on variable stars. The data from the AAVSO International Database (AAVSOID) are well calibrated and accurate. Currently, there are a total of 33,292 observations for \objectname{AM Herculis} recorded in the AAVSOID, and this number is always growing. Thus, it is necessary to take full advantage of the data in the AAVSOID. Based on this database, we have extracted 16 available optical light curves. Twelve light curves covering at least one primary minimum are listed in Figures 1(a)-(k), and the other four light curves with transient events are listed in Figure 2. The same parabolic fitting method and the method of estimating the timing uncertainty used in Section 2.1 were also applied to calculate the primary minimum times and the corresponding errors. The multi-band observations shown in Figure 1(c) clearly suggest a high consistency in the primary minimum timings of different optical bands.

\citet{kaf09} have pointed out that there are three subtypes in the high state of \objectname{AM Herculis} (i.e., a normal high state and another two newly defined high states labeled A and B in their Figure 1). Accordingly, we checked the photometric states of all 16 light curves in detail. Figures 1(c) and (d) show light curves in a failed high state and Figures 1(e)-(i) show data in a faint high state. All of the primary minimum times in three subtypes are included in our timing analysis due to the striking similarity of their light curves \citep{kaf09}. Finally, the 13 primary minimum times in the high state are measured from the AAVSOID. Since the timing system of the data used in the AAVSOID is JD, we have converted this timing system
to HJD for consistency with the literature.

\section{EPHEMERIS and O-C ANALYSIS}

All of the light curves shown in Figure 1 show that the primary minima of \objectname{AM Herculis} are stable in spite of the observed cycle-to-cycle changes in the profiles. The recent photometric ephemeris of \citet{kafa05},
\begin{equation}
T_{0}=HJD\ 244603.403(5)+0.12892704(1)\ E,
\end{equation}
is not only an updated ephemeris derived from high-state photometric data spanning more than 10 yr, but is also a spectroscopic ephemeris defining phase zero at the inferior conjunction of the secondary star. Compared with the old photometric ephemeris \citep{szk77}, the accumulated time error has grown up to ¡«11 minutes. In order to reduce the accumulated error in the timing analysis, we chose the recent ephemeris of \citet{kafa05} to calculate O-Cs for all 45 high-state light minimum times. Considering that the uncertainty of the epoch in Equation (1) is 0.005 days (i.e., $\sim$ 7.2 minutes), which can result in a large scatter in the O-C analysis, we revised this ephemeris and improved its precision based on the 45 high-state
data spanning 35 yr. Since there are 21 light minimum times published in the literature that lack corresponding uncertainties, we adopted a unified uncertainty, 8.8 days $\times10^{\rm -4}$, which was estimated from the arithmetic mean of the other 24 known errors. The corrections for the new epoch and orbital period of \objectname{AM Herculis} listed in Table 3 were calculated from a simple linear fit to the high-state O-Cs. Thus, a corrected photometric ephemeris, corresponding to the linear fitting case listed in Table 3, was derived as:
\begin{equation}
T_{0}=HJD\ 2446603.401658(99)+0.1289271368(17)\ E,
\end{equation}
with a variance of 4.3 days $\times10^{\rm -3}$. The difference in epoch between the new and old ephemerides is only $\sim$ 2 minutes. However, the precision of the new epoch has been improved by a factor of over 50. In addition, the other four nonlinear fitting formulae - linear-plus-sinusoidal, constant-plus-sinusoidal, quadratic-plus-sinusoidal, and quadratic-plus-LITE listed in Table 3 - are used to describe the O-C diagram of \objectname{AM Herculis} (see Figure 4), where the formulae refer to the solid lines, dotted lines, short lines, and long dashed lines, respectively. A Levenberg-Marquadt algorithm was used to carry out the three nonlinear least-squares fitting routines with strict sinusoidal term. The reduced $\chi^{\rm 2}$ in the three cases were all near unity, and the F-test proposed by \citet{pri75} confirms their high confidence levels, over 99\%. In order to further verify the existence of quadratic and periodic variations, we utilized an improved Nelder-Mead simplex algorithm, which is based on the famous nonlinear optimization method first proposed by \citet{nel65}. We used this algorithm to analyze the high-state O-C diagram. The fitting formulation is described by \citet{irw52}:
\begin{equation}
\left \{
\begin{array} {l}
O-C=\alpha+\beta E+\gamma E^{2}+\tau_{\rm LITE} \\
\\
\tau_{\rm LITE}=\frac{\kappa}{\sqrt{1-e^{2}\cos^{2}\Omega}}[\frac{1-e^{2}}{1+e\cos \nu}\sin(\nu+\Omega)+e\sin\Omega] \\
\\
\tan{\frac{\nu}{2}}=\sqrt{\frac{1+e}{1-e}}\tan\frac{EE}{2} \\
\\
M=EE-e\sin{EE} \\
\\
M=\frac{2\pi}{P_{3}}(t-T_{0}) \\
\end{array}
\right.
\end{equation}
where E is the cycle and $\kappa$, e, $\Omega$, $\nu$, M, EE, $T_{0}$, $P_{3}$ and t are the semi-amplitude, eccentricity, longitude of periastron, true anomaly, mean anomaly, eccentric anomaly, time of periastron passage, orbital period of the third body, and the minimum time, respectively. In all, there are eight parameters ($\alpha$, $\beta$, $\gamma$, $\kappa$, $e$, $\Omega$, $T_{0}$ and $P_{3}$) in Equation (3). We have performed three important parameter searches for e, $P_{3}$ and $\gamma$. In each iteration, the searched parameters are fixed in turn, but changed with a small given step. Within the reasonable parameter range, we obtained a reduced $\chi^{2}_{\rm\zeta}$ as a function of the different parameter values. As can be seen in Figure 3, the clear minimum values for the three searched parameters are e(min) $\sim$ 0.5, $P_{3}$(min) $\sim$ 5400 days, and $\gamma$(min) $\sim\rm-10^{-11}$, respectively. All three minimum values are consistent with the final optimized parameters listed in Table 3, to some extent. The output of the Nelder-Mead simplex algorithm was assigned to be an initial input parameter for the Levenberg-Marquadt algorithm for estimating the corresponding parameter errors. Note that Figure 3 illustrates some important facts: the eccentricity can
be acceptable as long as e $<$ 0.5, the optimized quadratic coefficient $\gamma$ is extremely close to zero, and it is hard to find the best cyclical period within a range of 5000-5400 days. Hence, the true errors of the three parameters may be larger than those listed in Table 3. It is interesting to note that the period and amplitude of the sinusoidal variation in all four cases are almost identical. Therefore, a periodic modulation with a period of 12-15 yr and an amplitude of 6-9 minutes in the O-C diagram of \objectname{AM Herculis} cannot be neglected. For the first time, a sinusoidal-like oscillation shown in Figure 4 is found for the prototypical polar \objectname{AM Herculis}. In Section 4.2, we will discuss the possible secular orbital period variation of \objectname{AM Herculis} in light of the above timing analysis.

\section{Discussion}

\subsection{Transient Events}

Low-state transient events are common phenomena in polars
\cite[e.g.][]{ram04,pan05,ara05}. For the prototypical polar \objectname{AM Herculis}, there are plenty of low-state transient events \citep[e.g.][]{sha93,bon00,kafb05,kal08,kaf09,ter10}. In this paper, seven new low-state transient events are detected. Figures 2(a)-(d) show the prominent events in the light curves acquired during the 2006 low state and the end of the high-to-low state transition in early 2008 August. Figures 2(e) and (f) show the events in the egress branches of light minima acquired during the beginning of the low-to-high state transition in early 2009 April. The two light curves in the 2006 low state lie right in the duration of the short flares labeled "C" in \citet{kaf09}. All of the light curves in Figure 2 phased by the ephemeris in Equation (2) show that the transient events are confined to a phase space between 0.0 and 0.5. Although all seven events spanning 2006 to 2009 were not observed in a series of sequential nights, every two light curves were obtained in the same season of the same year. From top to bottom, the seven events clearly demonstrate a distinct phase drift as a function of time. It is interesting that the direction of the phase drift in the three events of 2009 are opposite to the direction of the phase drifts of the previous four events. Additionally, the 9 mag excesses observed by \citet{kal08} were also scattered in orbital phase. Hence, the detected phase drift in Figure 2 may imply that the low-state events of \objectname{AM Herculis} likely have no phase dependence, which further supports the distribution of the flaring events in the long-term RoboScope data \citep{kafb05}.

Among the seven events, there are two special optical events with striking exponential profiles shown in Figures 2(d) and (e). These events resemble the X-ray flare event observed on 2008 October 30 by Suzaku \citep{ter10}. All three optical transient events were detected in a short-duration low state from 2008 August to 2009 April. The striking similarity in profile between optical and X-ray events may imply a similar driving mechanism. An offset exponential function similar to that used in the X-ray event \citep{ter10},
\begin{equation}
F=F_{0}+Ae^{-\frac{p-p_{0}}{\tau_{\rm flare}}},
\end{equation}
is used to describe the transient events in Figures 2(d) and (e), where F0, A, p, $p_{0}$ and $\tau_{\rm flare}$ are the magnitude of the post event, the amplitude of the event, the orbital phase, the initial orbital phase of the event, and the decay constant in phase units, respectively. The parameters for both events are listed in Table 4. Assuming that the end phase of the event corresponds to a magnitude difference of 0.1 mag between the event and the post event, the duration of the event $\Psi$ can be estimated. The decay constant $\tau_{\rm flare}$ for both optical events is less than for the X-ray flare event. This result may indicate that transient events for polars with similar exponential forms are high-energy events. According to Equations (2) and (3) in \citet{ter10}, the smaller ¦Óflare cannot yet be regarded as the cooling timescale of the thermal plasma. Considering that both optical events occurred near phase zero, in accordance with the phase of the X-ray flare event (i.e., $\sim$ 0.1), the transient events in exponential form may not only be caused by a same mechanism, but may also come from the same location.

On the other, there are five symmetric transient events (i.e., nearly equal rise and fall branches) shown in the other four light curves, which are different in shape from the V-band OPTIC flare during the 2004 low state, which had a faster rise branch than fall (Kafka et al. 2005b). A Gaussian function
\begin{equation}
F=F_{0}+Ae^{-(\frac{p-p_{0}}{ \Gamma})^2},
\end{equation}
where $\Gamma$ is equal to $\sqrt{2}\Psi$/6, is used to fit these transient events. Table 4 lists the details of the five events, including the lasting time, the peak phase, and the amplitude. The brightening events in Figures 2(a), (b), and (f) present typical amplitudes of 0.2-0.6 mag and durations of several tens of minutes \citep{bon00,kafa05,kafb05}. Although there is seemingly not any obvious transient event in the light curve observed on 2008 August 4, the violent variations in phase near $\sim$ 0.15 and $\sim$ 0.4 are distinctly alike the steep rise and fall branches of a symmetric transient event. Additionally, the light curve is flat outside the variations from $\sim$ 0.15 to $\sim$ 0.4, and seem to repeat this variation after $\sim$ 0.6. Therefore, we deduced the bold hypothesis that the low-state light curve of \objectname{AM Herculis} may hide a lot of periodic huge events, which are usually disturbed by unclear mechanisms, unfortunately. Assuming that the residual accretion in the low state of \objectname{AM Herculis} is blobby accretion \citep{lit90,gan95}, which caused the huge 1992 event \citep{bon00}, it may be possible that the occasional occultation of the accretion blobs results in the two dips in the "disrupted" plateaus of event. A Gaussian function is also used to fit its rise and fall branches, as with the other symmetric events. The best-fitting Gaussian curve denoted by the dashed line in Figure 2(c) seem to reappear in the "disrupted" huge V-band flare event with an amplitude of $\sim$ 2 mag and a duration of $\sim$ 72 minutes. It is interesting that this "disrupted event" is strikingly similar to the Rc-band flare event shown in Figure 4(b) of \citet{kal08}. However, the duration of $\sim$ 72 minutes is three times longer than that of the 1992 event \citep{sha93,bon00}. Since this light curve was observed just at the end of the high-to-low state transition, the whole system of \objectname{AM Herculis} may be undergoing a violent adjustment from the high state to the low state. Based on the starspot-covering model near the L1 point \citep{liv94}, this huge "disrupted event" may be an important manifestation of the strengthening of stellar magnetic activities on the surface of the red dwarf during the end of the high-to-low state transition.

On the other hand, the two continuing R-band brightening events with similar amplitudes and durations occurring in the same egress branch of the light minimum shown in Figure 2(f) were just detected at the beginning of the low-to-high state transition. Assuming that these continuing twin transient events were induced by the residual accretion located at the magnetic pole regions of the white dwarf, they may indicate that the two poles are active right now. If this assumption is true, then the observed twin events would be strong evidence for the conclusions of \citet{kaf09}: the overall high-state accretion geometry has already been established at the end of the low state. Since an explanation for the accretion events at the two poles cannot be confirmed completely at the present, the stellar activity on the secondary star should also be discussed. Most of the R-band flares in cool stars usually have low amplitudes of less than 0.15 mag, accompanied by occasional large amplitudes over 0.2 mag \citep[e.g.][]{zei82,koz04,koz06,nel07,vid09,zha10}. The largest R-band flare in the M-type eclipsing binary CU Cnc detected recently by \citet{qia12} only had an amplitude of 0.52 mag. Consequently,
the successive twin events of \objectname{AM Herculis} shown in Figure 2(f) may be two continuing large R-band flare events on a M4-5 type red dwarf \citep{you81,bai88,gan95,sou95}. This occurrence may be just a coincidence, but the eruptive prominences resulting from coronal mass ejections on the fast rotating secondary star have already been previously observed in several polars \citep[e.g.][]{sha85,sch93,pan02,pan05}. Does this result mean that the red dwarf in polars has been affected by the magnetic white dwarf? In addition, we cannot neglect another possible interpretation that both similar events originate from different physical regions in \objectname{AM Herculis} (i.e., one is from the magnetic pole and the other is from the flare). Hence, simultaneous X-ray and optical observations are necessary in the future for identifying all of these low-state transient events in \objectname{AM Herculis}.

Moreover, we noted that the two light curves observed in the 2009 low state likely show a significant modulation near phase zero. However, the top four low-state light curves in Figure 2 never present any modulation. In particular, the entire light curve in Figure 2(b) is almost flat, except for the event near phase 0.33. Since the two low-state light curves were observed at the beginning of the 2009 low-to-high state transition, their modulations may be caused by the recovered high-state accretion geometry. This result implies that the humps in Figures 2(e) and (f) near phase 0.7 with a typical amplitude of $\sim$ 0.6 mag and the extremely long phase duration of $\sim$ 0.4 (via $\sim$ 74 minutes) may be caused by the cyclotron emission region located at the accreting column(s). Our data alone cannot totally rule out the explanation of an accretion event or flare event on secondary star. In this case, \objectname{AM Herculis}, with its rich series of events/bursts, could be a highly active binary system, marked by the rapid establishment of a high-state accretion geometry at the low-to-high state transition.

Except for the brightening events in Figure 2, there is a special V-shaped light minimum event with a duration of $\sim$ 15 minutes and an amplitude of $\sim$ 0.7 mag shown in Figure 1(m) occurring at a phase $\sim$ 0.6. The amplitude of this minimum is almost the same as that of the primary minimum. However, this minimum is totally different from the secondary minimum of \objectname{AM Herculis}
\citep{szk77,szk80,gan01}. If this minimum is regarded as a peculiar secondary minimum, then the accretion region near the magnetic pole(s) of the white dwarf would be reduced to one-fifth of its normal size and deflected $\sim$ 30$^{\circ}$. away from its original location. This dramatic variation does not justify the stable high-state accretion configuration. Additionally, the two temporary plateaus near phase 0.7 and phase 0.9 with a duration of $\sim$ 9 minutes, denoted by the two dashed-dot circles in Figure 1(m), have also been detected clearly in the ingress branch of the primary minimum. Thus, it is more probable that the normal cyclotron beaming modulation cycle is disturbed coincidently by some uncertain short-duration occultations. Unfortunately, the lack of a light curve before this V-shaped minimum cannot be used to reveal more information about this peculiar minimum.

\subsection{Secular Orbital Period Variations}

Inspection of the lower panel of Figure 4 indicates that the two best-fitting curves seem to clearly deviate away from the data point at cycle 36254. In order to check the quadratic term, a simple quadratic function is used to fit this O-C diagram. The variance of the quadratic fitting, 4.7 days $\times10^{-3}$, is even larger than that of the linear fitting, 4.3 days $\times10^{-3}$. This result means that the significant level of the quadratic term is zero. Conservatively, the quadratic variation in \objectname{AM Herculis} is uncertain, based on the present 45 high-state photometric data points. However, since the orbital period of \objectname{AM Herculis} is located just at the upper edge of the period gap, further investigation may be necessary to understand the secular orbital period variations in \objectname{AM Herculis}.

Considering that the physical parameters of the donor star are key for the evolution of CVs \citep[see, e.g.,][]{rap83,spr83,how01}, we attempt to investigate its mass and radius derived by \citet{you81}. Corresponding to a normal main sequence star with a mass of $\sim$ 0.26 M$_{\odot}$, the radius in thermal equilibrium should be Re $\sim$ 0.27 R$_{\odot}$ based on Table 15.8 of \citet{cox00}. Hence,
the donor star of \objectname{AM Herculis}, with a radius of R2 $\sim$ 0.32 R$_{\odot}$, is obviously oversized by a bloating factor $f(R_{2}/R_{e})$ $\sim$ 1.2, which is just a lower limit corresponding to the shorter width (3/4 hr) of the period gap \citep{how01}. In addition, f can also be obtained from the mass and radius relations of the donor star as a function of the orbital period, as deduced by \citet{how01}:
\begin{equation}
\begin{array} {rl}
M_{2}(P_{orb})\simeq0.08f^{-1.95}P_{orb}^{1.3},\\
\\
R_{2}(P_{orb})\simeq0.10f^{-0.65}P_{orb}^{1.1},\\
\end{array}
\end{equation}
where M$_{2}$ and R$_{2}$ are in solar units and $P_{\rm orb}$ is in hours. In sum, the donor star of \objectname{AM Herculis}, with a mass of 0.26 M$_{\odot}$ and a radius of 0.32 R$_{\odot}$, is in accordance with a description of the conventional picture of CV evolution.

In light of the best-fitting parameters of the formulae with quadratic terms listed in Table 3, the orbital period decrease rate of \objectname{AM Herculis} derived from the average of two quadratic terms is $\dot{P}_{\rm orb}\sim-7.8(\pm1.2)\times10^{-11}$ s s$^{-1}$ (i.e., $\dot{P}_{\rm orb}/P_{\rm orb}\sim-7(\pm1)\times10^{-15}$ s$^{-1}$), which is one order of magnitude lower than the previous results for \objectname{AM Herculis} \citep{you79,maz86}. However, these results are the same order of magnitude as the results of other CVs, such as the Z-Cam type dwarf nova EM Cygni \citep{daia10}, an old post-nova T Aurigae \citep{daib10}, and so on. Since the masses of the white dwarf and the red dwarf in \objectname{AM Herculis} are 0.78 M$_{\odot}$ and 0.26 M$_{\odot}$, respectively \citep{you81}, conservation of mass transfer from the red dwarf to the massive white dwarf cannot result in the orbital period decrease. Therefore, there is no justification that the upward parabolic variation in the O-C diagram given by \citet{kal08} should be regarded as the consequence of conservative mass transfer. The updated O-C diagram shown in Figure 4 spanning the 30 yr data set has refuted the pseudo upward
variation. On the other hand, the average mass accretion rate of
$\sim$ 10$^{-10}$ M$_{\odot}$yr$^{-1}$ for both states \citep[e.g.,][]{you79,hes00,tow03,gan01,gan06}, which is small enough to be driven only by the gravitational radiation, suggests that the mass-loss timescale of the donor, $\tau_{\dot{\rm M}_{2}}\sim2.6\times10^{9}$ yr, is a little shorter than its Kelvin¨CHelmholtz timescale, $\tau_{\dot{\rm M}_{2}}\sim2.6\times10^{9}$ yr. Both timescales are comparable, which may imply that \objectname{AM Herculis} is indeed approaching the period gap, according to the standard evolutionary scenario for CVs \citep[see, e.g.,][]{spr83,rap83,how01,kni12}. Using the formula of \citet{pac71},
\begin{equation}
\frac{R_{\rm L_{2}}}{a}=0.462(\frac{M_{2}}{M})^{1/3},
\end{equation}
where a is the binary separation and M is the combined mass of
two component stars (i.e.,M = M$_{1}$+M$_{2}$), the Roche-lobe radius of the donor star can be calculated to be $R_{\rm L_{2}}\sim0.27$ R$_{\odot}$, which is smaller than the radius of the donor star $R_{2}\sim0.32$ R$_{\odot}$.Based on Kepler¡¯s third law, a logarithmic differentiation of Equation (7) yields:
\begin{equation}
\frac{\dot{R}_{\rm L_{2}}}{R_{\rm L_{2}}}=\frac{\dot{M}_{2}}{3M_{2}}+\frac{2}{3}\frac{\dot{P}_{\rm orb}}{P_{\rm orb}}.
\end{equation}
Hence, $\dot{R}_{\rm L_{2}}/R_{\rm L_{2}}\simeq-1.48\times10^{-7}$ yr$^{-1}$. Supposing that the variation of $R_{2}$ and $R_{\rm L_{2}}$ are synchronized (i.e., $\tau_{\dot{\rm R}_{2}}\simeq(\dot{R}_{\rm L_{2}}/R_{\rm L_{2}})^{-1}$), the average radius adjustment timescale of the donor star, $\tau_{\dot{\rm R}_{2}}$, can be estimated to be $6.8\times10^{6}$ yr. This result means that $\tau_{\dot{\rm R}_{2}}\simeq0.02\tau_{\rm KH}$. On the basis of the detailed reviews and the analytical estimates of $\tau_{\dot{\rm R}_{2}}$ for stars with a substantial convective envelope \citep{ste96,kni12}, this timescale amply satisfies the requirements of the sharp edges of the period gap and the well-defined cutoff at $P_{\rm min}$ in the CV distribution demonstrated by all numerical calculations of the evolution of CVs with different initial parameters \citep[see, e.g.,][]{pac83,kol92,how01}. Nevertheless, if we neglect the second term on the right side of Equation (8), then $\tau_{\dot{\rm R}_{2}}\sim7.8\times10^{9}$ yr would be longer than $\tau_{\rm KH}$, which contradicts the rapid convergence of CV evolution tracks
at $P_{\rm orb}\approx3$ hr \citep{how01}. Thus, it is necessary to seek possible mechanism for dissipating angular momentum in \objectname{AM Herculis}.

Assuming that the evolution of \objectname{AM Herculis} indeed causes the orbital period decrease with a rate similar to our observations, magnetic braking may be a possible explanation, in accordance with the standard theory of CV evolution \citep{rob81,spr83}. A consideration of the counteraction between the mass transfer and the gravitational radiation means that magnetic braking can account completely for the observed orbital period decrease, $\dot{P}_{\rm orb}\sim-7.8(\pm1.2)\times10^{-11}$ s s$^{-1}$, which corresponds to an angular momentum loss rate of $\dot{J}_{\rm orb}\sim-2.9\times10^{36}$ dyn cm. At first, we considered two magnetic braking models proposed by \citet{rap83} and \citet{web02}, which we denote as MB1 and MB2, respectively. The MB1 model is a common magnetic braking model for non-magnetic CVs with index parameter $\gamma_{\rm MB}=3$, while the MB2 model calculates a detailed magnetic interaction in the combined magnetospheres of two component stars for magnetic CVs. Taking advantage of the physical parameters of \objectname{AM Herculis} derived by \citet{you81}, the two models produce a similar angular momental loss rate, $\dot{J}_{\rm MB1}\sim-2.7\times10^{35}$ dyn cm and $\dot{J}_{\rm MB2}\sim-1.7\times10^{35}$ dyn cm, respectively, which are about one order of magnitude smaller than $\dot{J}_{\rm orb}$. As pointed out by \citet{wic94,li94,web02} and \citet{cum02}, the magnetic braking in \objectname{AM Herculis} should not be cut off totally, but should rather be strongly suppressed. This suppression is due to the magnetic moment of the white dwarf, ¦Ì$\mu_{1}\simeq6.2\times10^{33}$ G cm$^{3}$, corresponding to $B_{\rm WD}\sim$ 14 MG, which is smaller than the theoretical critical value, $\mu_{\rm crit}\simeq4\times10^{34}$ G cm$^{3}$, given by \citet{cum02}. Thus, magnetic braking in \object{AM Herculis} cannot be a viable angular momentum loss mechanism, at least not yet. Other than magnetic braking, three other mechanisms that direct mass loss from the binary system and the outflows from the Lagrangian point $L_{2}$ and the circumbinary disk, can also be involved to explain this large angular momentum loss rate. The first mechanism requires a high mass loss rate of $\sim-3.8\times10^{-8}$ M$_{\odot}$ yr$^{-1}$. The latter two models, which have been investigated by \citet{sha12} in particular, can be described by the following formulae:
\begin{equation}
\large
\begin{array} {l}
\frac{\dot{J}_{\rm L_{2}}}{J_{\rm orb}}=(\frac{a_{\rm L_{2}}}{a})^{2}\frac{M}{M_{1}}\frac{\dot{M}_{\rm L_{2}}}{M_{2}},\\
\\
\frac{\dot{J}_{\rm CB}}{J_{\rm orb}}=\gamma_{\rm CB}\frac{M}{M_{1}}\frac{\dot{M}_{\rm CB}}{M_{2}},\\
\end{array}
\end{equation}
where $a_{\rm L_{2}}$ is the distance between the mass center of the binary and the $L_{2}$ point and $\gamma_{\rm CB}$ is the fraction of the inner radius of the circumbinary disk. When we adopt the ratio $a_{\rm L_{2}}/a=1.51425$ with regard to the mass ratio $q\simeq0.33$ from Table 1 of \citet{moc84} and $\gamma_{\rm CB}=1.5$ from \citet{sob97}, we obtain mass-loss rates of $\dot{M}_{\rm L_{2}}\sim-6.3\times10^{-9}$ M$_{\odot}$ yr$^{-1}$ and $\dot{M}_{\rm CB}\sim-9.6\times10^{-9}$ M$_{\odot}$ yr$^{-1}$. Therefore, all three models require inconceivably high mass-loss rates.

Unless there is another unknown but more readily available angular momentum loss mechanism that forces the evolution of \objectname{AM Herculis}, this object will stay on the upper edge of the period gap for a long time and evolve slowly via gravitational radiation and severely curtailed magnetic braking. This result may support the argument that polars have their own special evolutionary track, different from the other non-magnetic CVs, as proposed by \citet{wic94,li94,web02}. If this stagnation in evolution is true and common for magnetic CVs with orbital periods of 3-4 hr, then could it be possible that the period gap is caused partly by this accumulation? On the other hand, \citet{kni12} proposed recently that the angular momentum loss rate for CVs below the gap is badly underestimated, and that the required angular momentum loss rate should be 2.47 times larger than that afforded by gravitational radiation. Thus, is it possible that the large $\dot{J}_{\rm orb}$ of \objectname{AM Herculis} calculated from its updated O-C diagram is caused by this mysterious angular momentum loss mechanism, since \objectname{AM Herculis} evolves into the upper edge of the gap?

\subsection{Periodic Modulation}

A prominent feature in the O-C diagram is the sinusoidal variation. Since the light minima photometric are attributed to the occultation of bright accretion spots by the limb of the white dwarf, a periodic spot motion may cause corresponding shifts in the light minimum times. In the eclipsing polar HU Aqr, \citet{sch01} claimed a change in the spot longitude of $\sim10^{\circ}$ between the different states.Moreover, \citet{cro81} reported that the inclination of the dipolar axis of \objectname{AM Herculis} has increased by $\sim5^{\circ}$ from 1976 to 1979. This finding may support the hypothesis of dipolar axis precession around the binary axis proposed by \citet{bai81}. However, the amplitude of the modulation shown in Figure 4, $\sim$ 0.005 days (i.e., $\sim$ 0.04 P$_{orb}$), indicates that the dipolar axis should shift nearly 14$^{\rm \circ}$. within 12-15 yr, which is larger than the conclusion of \citet{cro81}. Thus, further evidence is needed to support this hypothesis. Moreover, a dipolar axis precession with an amplitude of $5^{\circ}$-$10^{\circ}$ will only cause a small offset in the light minimum times, $\sim3$ minutes (i.e. $\sim$ 0.02 P$_{\rm orb}$). This result means that more high-precision photometry is needed to clarify this spot motion mechanism.

The modulations in the O-C diagrams of many other eclipsing polars such as UZ For \citep{pot11,daic10}, DP Leo \citep{beu11,qia10}, and HU Aqr \citep{goz12,qia11} are interpreted by the light travel-time effect, which is caused by a perturbation from a tertiary component. Since the optimized parameters of the sine-like modulation listed in Table 3 are similar for the four non-linear fitting cases, we chose a set of parameters of linear-plus-sinusoidal fitting to estimate the
projected distance from the binary to the mass center of a triple
system, $A_{12}\sin{i_{3}}\sim0.89(\pm0.02)$ AU, and the mass function of the third component, $f(M_{3})\sim4.5(\pm0.4)\times10^{-3}$ M$_{\odot}$. A combined mass of 0.78 M$_{\odot}$ + 0.26 M$_{\odot}$ is used to calculate the two relationships of $M_{3}$ versus $i_{3}$ and $A_{3}$ versus $i_{3}$. Both relationships shown in Figure 5 clearly indicate that this assumed third component should be a red dwarf with a mass similar to the secondary star, as all of them are in a coplanar orbit. Its mass is close to 0.2 M$_{\odot}$, even if $i_{3}$ reaches up to $90^{\circ}$. The separation between the third body and the binary is three orders of magnitudes larger than that of the binary itself, $a\sim0.92$ R$_{\odot}$, as long as the orbital inclination is moderate (i.e., $i_{3}>30^{\circ}$). In an eccentric orbit, $M_{3}$ and $A_{3}$ can be calculated using the following formulae \citep{irw52,may90}:
\begin{equation}
\begin{array} {l}
f(M_{3})=\frac{(A_{12}\sin{i_{3}})^{3}}{P^{2}_{3}}=\frac{1}{P^{2}_{3}}(\frac{173.15\kappa}{\sqrt{1-e^{2}\cos^{2}\Omega}})^{3},\\
\\
A_{3}=A_{12}\frac{M_{1}+M_{2}}{M_{3}},\\
\end{array}
\end{equation}
where $A_{12}$ and $P_{3}$ are in AU and years, respectively. Although the significance level and the parameter errors listed in Table 3 suggest that quadratic-plus-LITE fitting is not an optimized description of the O-C diagram of \objectname{AM Herculis} compared with the other three cases with strict sinusoidal terms, Figure 5 obviously shows that the thin lines denoting the eccentric orbit are close to the other thick lines referring to the circular orbit, except for the small difference between the two dashed lines at high inclinations. Hence, to some extent, the eccentricity $\sim$ 0.47 never changes the dynamic feature of the third star seriously. Compared with the gravitational binding energy of the two components of \objectname{AM Herculis}, the interaction between the third red dwarf and the binary system can be neglected, which means that a dynamic stable triple system can be expected. If this assumed a red dwarf indeed exists, then \objectname{AM Herculis} would be a peculiar system with an amazing configuration and its formation and evolution would be worthy of further investigation. In fact, it is hard to definitely distinguish the secondary and third components in observations since their discrepancies in the coplanar case are only in their spin velocity and reflection effect. On the other hand, the abundance of incredible and complicated transient events in \objectname{AM Herculis} may badly blur the two red dwarfs. In particular, compared with the brilliant accreting magnetic white dwarf, the observational features of the two red dwarfs would be too faint to be detected. This may be why this massive third component is always invisible in all kinds of observations.

On the other hand, the periodic changes in the O-C diagram of \objectname{AM Herculis} may reflect a true orbital period variation because the period measured from the photometric light curves in optical bands is nearly equal to \objectname{AM Herculis} real orbital period. At present, plenty of investigations of the significant discrepancy between theoretical and observational mass-radius relations for low-mass M-type dwarfs have found that there is strong magnetic activity in the components of short-period red-dwarf binaries \citep[e.g.][]{qia12,tor10,mor10,dev08}. Furthermore, a lot of photometric flare events of \objectname{AM Herculis} in the low state have been detected in this paper and the literature \citep[e.g.][]{ter10,kal08,kafb05,bon00}. For this reason, there may be a strong solar-type magnetic activity cycle in the M4-5 type secondary star of \objectname{AM Herculis}. According to Applegate¡¯s mechanism \citep{hal89,app92}, the variation of the quadrupole momentum $\triangle Q$ for the secondary star (i.e., a fully convective dwarf), can be calculated to be $\sim-1.7(\pm0.9)\times10^{48}$ g cm$^{2}$. Then, a relationship between the minimum energy required to drive the observed period oscillation and the assumed shell mass of the secondary star shown in Figure 6 indicates that an M4-5 type red dwarf cannot afford sufficient energy to force the periodic modulation observed in the O-C diagram. Additionally, based on an extended model with Applegate¡¯s considerations proposed by \citet{lan05,lan06}, the longest timescale of energy dissipation $\tau\approx0.8$ yr for the magnetic activity in \objectname{AM Herculis} was estimated from the eigenvalue of the equation of angular momentum conservation, $\lambda_{20}\approx3.8\times10^{-8}$ s$^{-1}$. Considering that the secondary star of \objectname{AM Herculis} with a high rotation speed is locked in a synchronous system with a strong magnetic white dwarf, it may not be suitable for this extended model proposed by \citet{lan06} due to its somewhat abnormal stellar structure. Moreover, the modulation period $P_{\rm mod}\sim P_{3}$, and the angular velocity of the secondary star, $\Omega_{\rm orb}\sim5.6\times10^{-4}$ rad s$^{-1}$, agree well with the regression relationship deduced by \citet{lan99}:
\begin{equation}
Log P_{\rm mod}=-0.36(\pm0.10)Log \Omega_{\rm orb}+0.018.
\end{equation}
Since \objectname{AM Herculis} is a prototype in the category of polars \citep[see][chap.6]{war95}, the secondary stellar activity in AM-Her-type binaries may be a common phenomenon due to the similar physical and geometric structures. Recently, many studies of the emission-line spectra of MCVs, for example, VV Pup \citep{howa06,mas08}, ST LMi \citep{kaf07} and EF Eri \citep{howb06}, have provided much evidence for chromospheric activity on the secondary star. However, there is no denying the fact that the large uncertainty on the slope in Equation (11) implies that $P_{\rm mod}$ in the range of 7-33 yr can be equally well accepted. The evidence weakens the hypothesis that the magnetic activity cycle in the secondary star causes the quasi-periodic modulations observed in the O-C diagram of AM Her-type binaries. An updated high-precision
logarithmic relationship of $P_{\rm mod}$ versus $\Omega_{\rm orb}$ is necessary for future studies.

\section{CONCLUSIONS}

From our 2009 photometry and the AAVSOID, we have obtained 21 light curves covering the low and high states shown in Figures 1 and 2. There are 16 new primary light minimum times in the high state, including 13 data points from the AAVSOID and 3 data points from our photometry. Furthermore, all the light minimum times published in the literature have been reexamined in detail. In view of the detailed classification of the high state of \objectname{AM Herculis} described by \citet{kaf09}, there are two data points in failed high state and six data points in a faint high state. All 45 high-state light minimum times are used to revise the photometric ephemeris of \citet{kafa05}. The revised ephemeris, now with high precision, has been used for the timing analysis and the calculation of orbital phase.

The phased seven low-state transient events spanning from 2006 to 2009 show obvious phase drifts and a weak phase dependence. Theses findings may be strong evidence for the results of \citet{kafb05}. In the short-duration low state from 2008 August to 2009 April, the two optical events with typical amplitudes and durations have a striking exponential form similar to that of the X-ray flare event observed in the same low state \citep{ter10}. Except for the two events with exponential forms, the other five symmetric events are described by a standard Gaussian function. Among these events, a "disrupted flare event" with an amplitude of $\sim$ 2 mag and a duration of $\sim$ 72 minutes may hide in the light curves during the end of the high-to-low state transition. Additionally, at the beginning of the low-to-high state transition, continuing twin events with typical amplitudes of 0.6-0.7 mag are thought to be two huge R-band flare events comparable with the largest, newly found R-band flare with an amplitude of 0.52 mag in the M-type eclipsing binary CU Cnc \citep{qia12}. All of the peculiar transient events detected during state transitions may suggest that the accretion configuration of \objectname{AM Herculis} may be undergoing fierce and fast adjustments between the two states.

Since the orbital period of \objectname{AM Herculis} is located on the upper edge of the gap exactly, it is necessary to investigate its secular orbital period variations in detail. Before this investigation, we verified that the mass and radius of its donor star, as derived by \citet{you81} are in agreement with the description of the standard CV evolutionary theory \citep[see, e.g.,][]{rob81,pac83,spr83,how01}. A calculation indicates that the three timescales of mass loss,Kelvin¨CHelmholtz, and the average radius adjustment for the donor star satisfy the relation $\tau_{\rm\dot{M}_{2}}\simeq\tau_{\rm KH}>>\tau_{\rm\dot{R}_{2}}$, which not only definitely requires appropriate angular momentum loss
mechanisms, but also supports the results of many numerical calculations of CV evolution with different initial parameters when approaching the upper edge of the period gap \citep[see, e.g.,][]{pac83,kol92,how01}. Nevertheless, we cannot find any available mechanism with high enough angular momentum loss rates
on the basis of the three direct mass-loss models and the two magnetic braking models proposed by \citet{rap83,li94} and \citet{web02}.

Based on 45 high-state data, the O-C diagram of \objectname{AM Herculis} indicates significant sine-like variations with a period of 12-15 yr and an amplitude of 6-9 minutes. We attempted to apply three mechanisms - spot motion, the light travel time effect, and magnetic activity cycles in secondary star - to explain this O-C modulation. At first, the spot motion mechanism requires an unsupported dipolar axis precession with a large amplitude of $\sim14^{\rm\circ}$ and a duration of 12-15 yr. Then, the light travel-time effect indicates that a red dwarf with a similar mass to the secondary star in the coplanar case may be the assumed third component of \objectname{AM Herculis}. If this third body indeed exists in \objectname{AM Herculis}, it would be the first detection of a fantastic celestial system in which the secondary and tertiary components have nearly identical masses. In considering the number of low-state flare events in \objectname{AM Herculis} and recent investigations of strong magnetic activity in low-mass M-type dwarfs, it is important to discuss the magnetic activity cycles in the secondary star in \objectname{AM Herculis}. Although the secondary star cannot supply enough energy to drive the observed periodical changes, the modulation period $P_{\rm mod}$ and the angular velocity of the secondary star $\Omega_{\rm orb}$ in \objectname{AM Herculis} are in good agreement with the low-precision regression relationship
of $\Omega_{\rm orb}$ versus $\Omega_{\rm orb}$ derived by \citet{lan99}. In the future, other more convincing evidence of the three mechanisms is needed. Finally, the periodic variations shown in Figure 4 should be investigated further with high-precision photometry.

\acknowledgments

This work was partly supported by the West Light Foundation of The Chinese Academy of Sciences and the Chinese Natural Science Foundation (Nos. 11003040 and 111330007). The CCD photometric observations of AM Herculis were obtained with the 0.6 m and 1.0 m telescopes at Yunnan Observatory and the 0.85 m telescopes at Xinglong Observing Station of National Astronomical Observatories. We acknowledge with thanks the variable star observations from the AAVSO International Database, contributed by observers worldwide. The data used in this research were contributed by the following observers: Massimiliano Martignoni, Lewis M. Cook, Stephen E. Robinson, William Goff, J.E. Morgan, Bart Staels, Francois M. Teyssier, Keith A. Graham, David Jarkins, and Massimo Banfi. We are grateful to the Director of the AAVSO, Dr. Arne Henden, and the Science Director of the AAVSO, Dr. Matthew Templeton, for their comments on this paper.

\begin{figure}
\centering
\includegraphics[width=12.0cm]{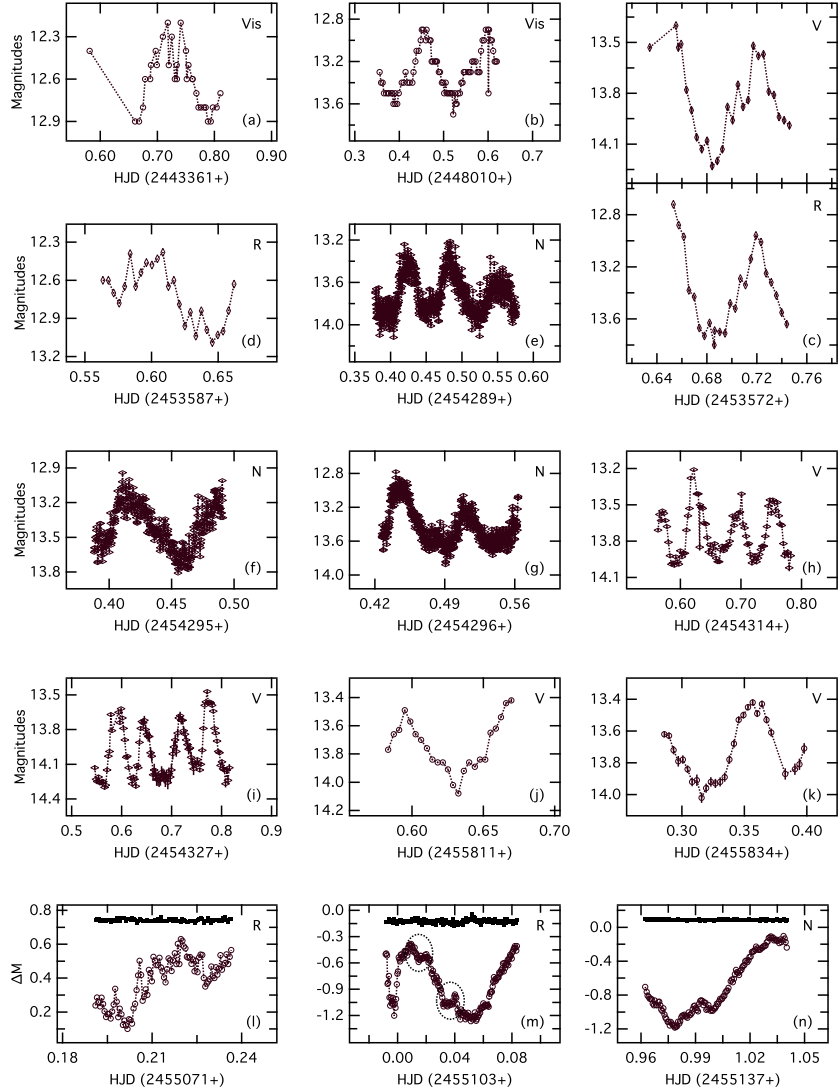}
\caption{Fifteen high-state light curves of the polar AM Herculis spanning 1977 to 2011. The light curves in the normal, failed, and faint high states are marked by the open circles, vertical diamonds, and horizontal diamonds, respectively. The photometric filter used is given in the top right of each plot. N and V refer to white light and visual observations, respectively.} \label{Fig. 1}
\end{figure}

\begin{figure}
\centering
\includegraphics[width=12.0cm]{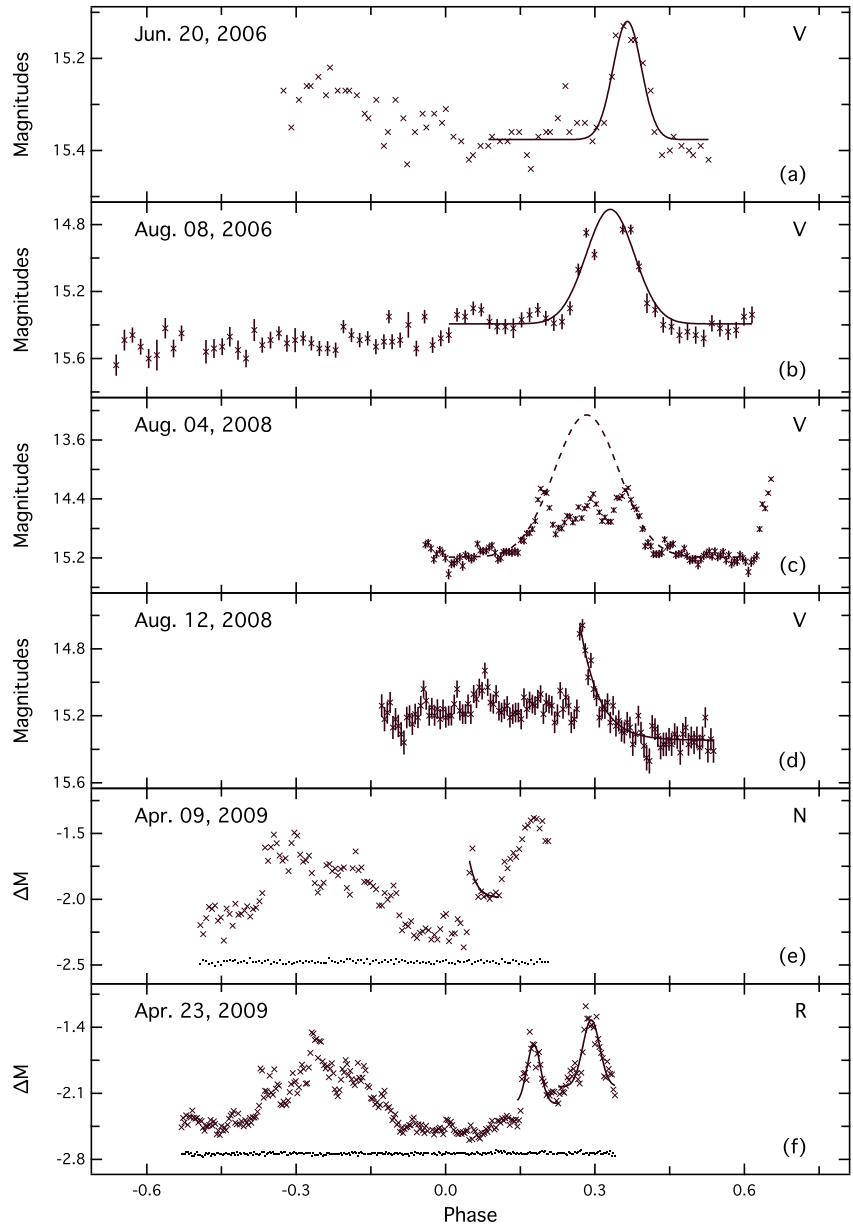}
\caption{Six low-state light curves with seven transient events are shown in time order from top to bottom. The solid lines denote the best-fitting curves for the really complete events. The dashed line in plot (c) denotes an assumed "disrupted" huge V-band flare event.} \label{Fig. 2}
\end{figure}

\begin{figure}
\centering
\includegraphics[width=9.0cm]{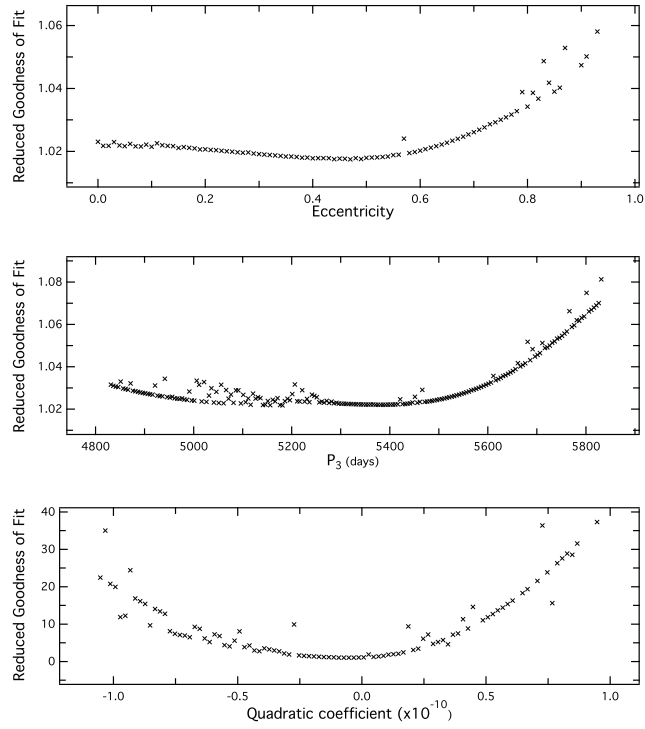}
\caption{Parameter searches for e, $P_{3}$ and $\gamma$ are displayed from top to bottom. The search ranges for the three parameters are $0\sim0.93$, $4831\sim5831$ days, and $-1\sim1(\times10^{-10})$, respectively.} \label{Fig. 3}
\end{figure}

\begin{figure}
\centering
\includegraphics[width=9.0cm]{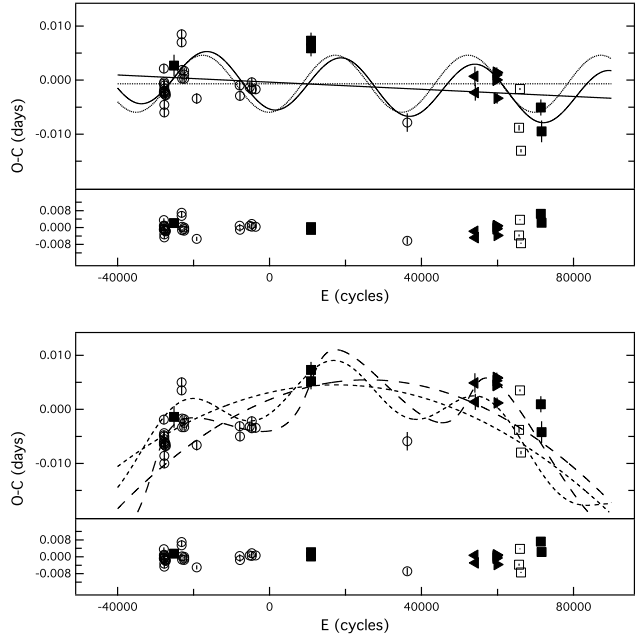}
\caption{Upper panel shows the constant-plus-sinusoidal and linear-plus-sinusoidal fitting curves, while the lower panel shows the quadratic-plus-sinusoidal and quadratic-plus-LITE curves. Open circles, open squares, solid squares, and leftward and rightward solid triangles refer to the normal high-state data from the literature, our own photometry, the AAVSO data, and the data in the failed and faint high states, respectively.} \label{Fig. 4}
\end{figure}

\begin{figure}
\centering
\includegraphics[width=9.0cm]{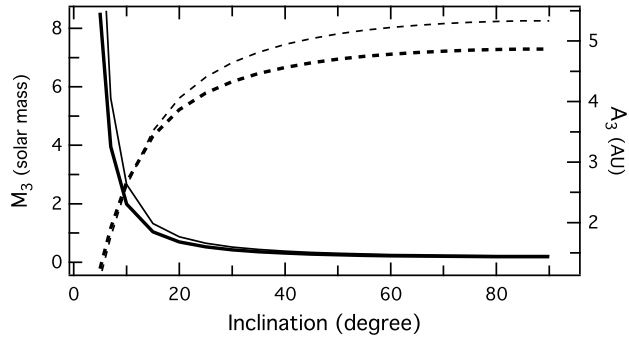}
\caption{Masses and separations of the third component in AM Herculis depending on its different orbital inclination are plotted with solid and dashed lines, corresponding to the left and right vertical axes, respectively.} \label{Fig. 5}
\end{figure}

\begin{table}
\begin{center}
\caption{Details of the Photometry.}
\begin{tabular}{cccccccc}
\tableline\tableline
No. & Date & UT$_{\rm start}$ & Telescope & Instrument & Filter & ET\tablenotemark{a} & Durations\tablenotemark{b}\\
&&& m & CCD && s &\\
\tableline
1. & 2009 Apr 09 & 18:10 & 1.0 & Andor DW436 2K & N\tablenotemark{c} & 60 & 0.70\\
2. & 2009 Apr 23 & 19:19 & 0.6 & Andor DW436 2K & R & 25 & 0.92\\
3. & 2009 Aug 27 & 16:33 & 0.85 & PI MicroMAX 1024 BFT & R & 30 & 0.36\\
4. & 2009 Sep 28 & 11:47 & 0.6 & Andor DW436 2K & R & 30 & 0.71\\
5. & 2009 Nov 02 & 11:06 & 1.0 & Andor DW436 2K & N\tablenotemark{c} & 40 & 0.61\\
\tableline\tableline
\end{tabular}
\tablecomments{}
\tablenotetext{a}{Exposure times}
\tablenotetext{b}{In phase units}
\tablenotetext{c}{Unfiltered photometry}
\end{center}
\end{table}

\begin{deluxetable}{p{3cm}cccccrc}
\tablewidth{0pt}
\tablecaption{Minimum Light Timings for the Polar AM Herculis.}
\tablehead{\colhead{JD.Hel.} & \colhead{errors} & \colhead{type} & \colhead{method} & \colhead{wavelength} & \colhead{E (cycle)} & \colhead{$(O-C)^{\rm d}$} & \colhead{Ref.}\\
\colhead{2400000+}&&&&&&&}
\startdata
43014.712660 &  & pri & pe  & V         & -27835 & -.00215 & (1)\\
43014.841268 &  & pri & pe  & V         & -27834 & -.00246 & (1)\\
43015.745535 &  & pri & pe  & V         & -27827 & -.00069 & (1)\\
43015.877313 &  & pri & pe  & V         & -27826 & +.00216 & (1)\\
43026.4452 &  & pri & pe  & V         & -27744 & -.00197 & (2,3)\\
43028.3807\tablenotemark{\star} & .0003  & pri & pe  & V\&R      & -27729 & -.00038 & (2,11)\\
43031.728621 &  & pri & pe  & V         & -27703 & -.00457 & (1)\\
43031.860554 &  & pri & pe  & V         & -27702 & -.00156 & (1)\\
43032.6347\tablenotemark{\star} &  & pri & pe  & y\&I         & -27696 & -.00098 & (4)\\
43033.6610 &  & pri & pe  & I         & -27688 & -.00609 & (4)\\
43062.5441\tablenotemark{\star} &  & pri & pe  & I\&FI         & -27464 & -.00267 & (4)\\
43062.802404 &  & pri & pe  & V         & -27462 & -.00222 & (1)\\
43069.6354 &  & pri & pe  & I         & -27409 & -.00237 & (4)\\
43083.5591 &  & pri & pe  & I         & -27301 & -.00280 & (4)\\
43361.789 & .002   & pri & Vis & Vis       & -25143 & +.00264 & (5)\\
43616.81305 &  & pri & pe  & V         & -23165 & +.00852 & (3,6)\\
43616.94045 &  & pri & pe  & V         & -23164 & +.00699 & (3,6)\\
43635.88762 & .00060  & pri & pe  & 690-740nm & -23017 & +.00187 & (7)\\
43636.78853 & .00060  & pri & pe  & 820-870nm & -23010 & +.00029 & (7)\\
43704.73386 & .00060  & pri & pe  & 800-835nm & -22483 & +.00102 & (7)\\
43704.86349 & .00060  & pri & pe  & 800-835nm & -22482 & +.00172 & (7)\\
43705.89342 & .00059 & pri & pe  & 750-780nm & -22474 & +.00023 & (7)\\
44133.0254 & .0010   & pri & pe  & V         & -19161 & -.00339 & (8,9)\\
45591.32259 &  & pri & pe  & V         &  -7850 & -.00104 & (3)\\
45600.34559 &  & pri & pe  & V         &  -7780 & -.00294 & (3)\\
45928.46670 &  & pri & pe  & V         &  -5235 & -.00140 & (3)\\
46000.27880 &  & pri & pe  & V         &  -4678 & -.00171 & (3)\\
46001.31151 &  & pri & pe  & V         &  -4670 & -.00042 & (3)\\
46132.55800 &  & pri & pe  & V         &  -3652 & -.00175 & (3)\\
48010.3894 & .0015  & pri & Vis & Vis       &  10913 & +.00590 & (5)\\
48010.5197 & .0015  & pri & Vis & Vis       &  10914 & +.00728 & (5)\\
51277.5182 & .0017  & pri & ccd & Unknown   &  36254 & -.00788 & (10)\\
53572.6876\tablenotemark{HF\star} & .0018  & pri & ccd & V\&R      &  54056 & +.00067 & (5)\\
53587.6402\tablenotemark{HF} & .0015  & pri & ccd & R        &  54172 & -.00231 & (5)\\
54289.3938\tablenotemark{Hf} & .0001  & pri & ccd & N/A       &  59615 & +.00092 & (5)\\
54289.5218\tablenotemark{Hf} & .0001  & pri & ccd & N/A       &  59616 & -.00005 & (5)\\
54295.4540\tablenotemark{Hf} & .0001  & pri & ccd & N/A       &  59662 & +.00151 & (5)\\
54296.4839\tablenotemark{Hf} & .0001  & pri & ccd & N/A       &  59670 & +.00008 & (5)\\
54314.664\tablenotemark{Hf} & .001   & pri & ccd & V         &  59811 & +.00137 & (5)\\
54327.681\tablenotemark{Hf} & .001   & pri & ccd & V         &  59912 & -.00336 & (5)\\
55071.1983 & .0005  & pri & ccd & R         &  65679 & -.00878 & (11)\\
55103.0485 & .0002  & pri & ccd & R         &  65926 & -.00161 & (11)\\
55137.9796 & .0003  & pri & ccd & N         &  66197 & -.01308 & (11)\\
55811.6306 & .0015  & pri & ccd & V         &  71422 & -.00506 & (5)\\
55834.317 & .002   & pri & ccd & V         &  71598 & -.00947 & (5)\\
\enddata
\tablecomments{}
\tablenotetext{\star}{The average data.}
\tablenotetext{HF}{The data in failed high state.}
\tablenotetext{Hf}{The data in faint high state.}
\tablerefs{(1) \citet{szk77}; (2) \citet{bai77}; (3) \citet{maz86}; (4) \citet{ols77}; (5)the AAVSO; (6) \citet{szk80}; (7) \citet{you79}; (8) \citet{bai81}; (9) \citet{you81};  (10) \citet{saf02};  (11) This paper.}
\end{deluxetable}

\begin{table}
\begin{center}
\caption{Least-squares Fittings for the O-C Diagram of AM Herculis.}
\begin{tabular}{llr}
\tableline\tableline
\textbf{Linear fitting}                      & $\alpha=-1.34(\pm0.10)\times10^{-3}$    & $\chi^{2}_{\rm\zeta}=70.8$\\
$O-C=\alpha+\beta E$                      & $\beta=9.68(\pm0.17)\times10^{-8}$   & $\zeta=43$\\
                                                               &                                                                & $\sigma=0^{\rm d}.0043$\\
\tableline
\textbf{Constant-plus-sinusoidal fitting}                      & $\alpha=-6.80(\pm2.30)\times10^{-4}$    & $\chi^{2}_{\rm\zeta}=39.1$\\
$O-C=\alpha+\kappa\sin[\omega E+\varphi]$                      & $\kappa=5.28(\pm0.16)\times10^{-3}$   & $\zeta=41$\\
                                                               & $\omega=1.80(\pm0.01)\times10^{-4}$     & $\sigma=0^{\rm d}.0034$\\
                                                               & $\varphi=4.74(\pm0.03)$                 & $\lambda_{\rm F-test}=12.98$\\
                                                               &                                             & Significant level: 100\%\\
\tableline
\textbf{Linear-plus-sinusoidal fitting}                        & $\alpha=-3.80(\pm2.50)\times10^{-4}$    &\\
$O-C=\alpha+\beta E+\kappa\sin[\omega E+\varphi]$              & $\beta=-3.31(\pm0.79)\times10^{-8}$   & $\chi^{2}_{\rm\zeta}=39.0$\\
                                                               & $\kappa=5.12(\pm0.14)\times10^{-3}$   & $\zeta=40$\\
                                                               & $\omega=1.78(\pm0.02)\times10^{-4}$     & $\sigma=0^{\rm d}.0034$\\
                                                               & $\varphi=4.48(\pm0.06)$                 & $\lambda_{\rm F-test}=8.57$\\
                                                               &                                             & Significant level: 99.98\%\\
\tableline
\textbf{Quadratic-plus-sinusoidal fitting}                     & $\alpha=3.12(\pm0.52)\times10^{-3}$   &\\
$O-C=\alpha+\beta E+\gamma E^{2}+\kappa\sin[\omega E+\varphi]$ & $\beta=1.60(\pm0.19)\times10^{-7}$    &\\
                                                               & $\gamma=-4.56(\pm0.64)\times10^{-12}$ & $\chi^{2}_{\rm\zeta}=37.5$\\
                                                               & $\kappa=4.52(\pm0.38)\times10^{-3}$   & $\zeta=39$\\
                                                               & $\omega=1.56(\pm0.04)\times10^{-4}$     & $\sigma=0^{\rm d}.0032$\\
                                                               & $\varphi=5.22(\pm0.07)$                 & $\lambda_{\rm F-test}=8.38$\\
                                                               &                                             & Significant level: 99.99\%\\
\tableline
\textbf{Quadratic-plus-LITE fitting}                     & $\alpha=1.80(\pm0.73)\times10^{-3}$      &\\
$O-C=\alpha+\beta E+\gamma E^{2}+\tau_{\rm LITE}$ & $\beta=2.84(\pm0.30)\times10^{-7}$      &\\
                                                               & $\gamma=-5.51(\pm0.84)\times10^{-12}$      &\\
                                                               & $\kappa=5.98(\pm2.49)\times10^{-3}$      &\\
                                                               & $P_{3}=5374^{\rm d}.04(\pm68.07)$      & $\chi^{2}_{\rm\zeta}=37.6$\\
                                                               & $e=0.47(\pm0.21)$      & $\zeta=37$\\
                                                               & $\Omega=0.28(\pm0.55)$      & $\sigma=0^{\rm d}.0033$\\
                                                               & $T_{0}=2448087.4(\pm300.0)$      & $\lambda_{\rm F-test}=1.96$\\
                                                               &                                             & Significant level: 90.39\%\\
\tableline\tableline
\end{tabular}
\end{center}
\end{table}

\begin{figure}
\centering
\includegraphics[width=9.0cm]{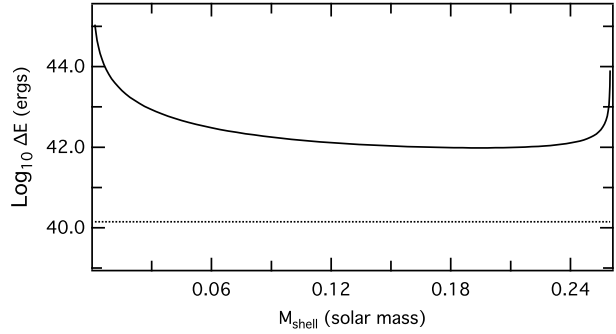}
\caption{Required energy for driving the observed cyclical oscillation in the O-C diagram of AM Herculis, corresponding to different shell masses in the secondary star, are plotted as a solid line. The dotted line refers to the total radiant energy of a M5-type red dwarf over the whole oscillation period of $\sim$ 12.5 yr.} \label{Fig. 6}
\end{figure}

\begin{table}
\begin{center}
\caption{Parameters of Low-state transient Events.}
\begin{tabular}{cccccccc}
\tableline\tableline
No. & Date & $F_{\rm 0}$ & A & $p_{\rm 0}$\tablenotemark{a} & $\Gamma$\tablenotemark{a} & $\tau_{\rm flare}$\tablenotemark{a} & $\Psi$\\
&& mag & mag & phase & phase & phase & min\\
\tableline
1. & 2006 Jun 20 & 15.376(7) & 0.26(2) & 0.365(3)  & 0.040(4) & - & 32(3)\\
2. & 2006 Aug 08 & 15.393(9) & 0.68(3) & 0.331(2)  & 0.069(3) & - & 54(3)\\
3. & 2008 Aug 04 & 15.189(9) & 1.93(7) & 0.2833(5) & 0.092(2) & - & 72(2)\\
4. & 2008 Aug 12 & 15.35(1)  & 0.70(3) & 0.269     & - & 0.039(4)  & 13(1)\\
5. & 2009 Apr 09 & -1.99(6)  & 0.28(8) & 0.048     & - & 0.015(11) & 3(2)\\
6. & 2009 Apr 23 & -2.21(6)  & 0.63(7) & 0.177(1)  & 0.019(3) & - & 15(2)\\
7. & 2009 Apr 23 & -2.03(4)  & 0.71(5) & 0.293(1)  & 0.024(3) & - & 19(2)\\
\tableline\tableline
\end{tabular}
\
\tablecomments{The error is in parentheses.}
\tablenotetext{a}{In phase units.}
\end{center}
\end{table}

\end{document}